\documentclass[twocolumn]{aastex63}
\usepackage{natbib,xcolor}
\usepackage[caption=false]{subfig}    % don't load caption or captcont
\shorttitle{Stellar Mass Fractions and Luminosity Functions in $z\sim1$ MaDCoWS Clusters}
\shortauthors{Decker et al.}

\newcommand{\eg}{e.g.,}

\newcommand{\ie}{i.e.,}
\newcommand{\spitzer}{{\it Spitzer}}

\newcommand{\msol}{M_\odot}

\newcommand{\madcows}{MaDCoWS}
\newcommand{\mfh}{$M_{500}$}
\newcommand{\rfh}{$r_{500}$}

\newcommand{\mstar}{m^*}

\newcommand{\cho}{$3.6~\mathrm{\mu m}$}
\newcommand{\cht}{$4.5~\mathrm{\mu m}$}
\newcommand{\fstar}{$f_\star$}

\newcommand{\Ncl}{12}
\newcommand{\mstarall}{19.83\pm0.12}
\newcommand{\mstarhigh}{19.86\pm0.20}
\newcommand{\mstarlow}{19.69\pm0.22}
\newcommand{\alphaall}{-0.81\pm0.10}
\newcommand{\alphahigh}{-0.95\pm0.15}
\newcommand{\alphalow}{-0.77\pm0.16}
\newcommand{\MHall}{-23.91\pm0.12}
\newcommand{\MHhigh}{-24.09\pm0.20}
\newcommand{\MHlow}{-23.80\pm0.22}

\begin{document}
\title{The Massive and Distant Clusters of WISE Survey XI: Stellar
Mass Fractions and Luminosity Functions of MaDCoWS Clusters at $z \sim
1$}
\author{Bandon Decker}
\affiliation{Department of Physics and Astronomy, University of
Missouri, 5110 Rockhill Road, Kansas City, MO 64110, USA}
\author{Mark Brodwin}
\affiliation{Department of Physics and Astronomy, University of
Missouri, 5110 Rockhill Road, Kansas City, MO 64110, USA}
\author{Ripon Saha}
\affiliation{Department of Physics and Astronomy, University of
Missouri, 5110 Rockhill Road, Kansas City, MO 64110, USA}
\author{Thomas Connor}
\affiliation{Jet Propulsion Laboratory, California Institute of
  Technology, 4800 Oak Grove Drive, Pasadena, CA 91109, USA}
\author{Peter R. M. Eisenhardt}
\affiliation{Jet Propulsion Laboratory, California Institute of
  Technology, 4800 Oak Grove Drive, Pasadena, CA 91109, USA}
\author{Anthony H. Gonzalez}
\affiliation{Department of Astronomy, University of Florida, 211
  Bryant Space Center, Gainesville, FL 32611, USA}
\author{Emily Moravec}
\affiliation{Astronomical Institute of the Czech Academy of Sciences,
Bo\v cn\'i ll 1401/1A, 14100 Praha 4, Czech Republic}
\author{Mustafa Muhibullah}
\affiliation{Department of Physics and Astronomy, University of
Missouri, 5110 Rockhill Road, Kansas City, MO 64110, USA}
\author{S. Adam Stanford}
\affiliation{Department of Physics, University of California, One
  Shields Avenue, Davis, CA 95616, USA}
\author{Daniel Stern}
\affiliation{Jet Propulsion Laboratory, California Institute of
  Technology, 4800 Oak Grove Drive, Pasadena, CA 91109, USA}
\author{Khunanon Thongkham}
\affiliation{Department of Astronomy, University of Florida, 211
  Bryant Space Center, Gainesville, FL 32611, USA}
\author{Dominika Wylezalek}
\affiliation {Astronomisches Rechen-Institut, Zentrum f\"{u}r Astronomie der Universit\"{a}t Heidelberg, M\"{o}nchhofstr. 12-14, D-69120 Heidelberg, Germany}
\author{Simon R. Dicker}
\affiliation{Department of Physics and Astronomy, University of
  Pennsylvania, 209 South 33rd Street, Philadelphia, PA 19104, USA}
\author{Brian Mason}
\affiliation{National Radio Astronomy Observatory, 520 Edgemont Road,
  Charlottesville VA 22903, USA}
\author{Tony Mroczkowski}
\affiliation{European Southern Observatory, Karl Schwarzschild Stra\ss
  e 2, D-85748, Garching bei M\"{u}nchen, Germany}
\author{Charles E. Romero}
\affiliation{Department of Physics and Astronomy, University of
  Pennsylvania, 209 South 33rd Street, Philadelphia, PA 19104, USA}
\affiliation{Green Bank Observatory, Green Bank, WV 24944, USA}
\author{Florian Ruppin}
\affiliation {Kavli Institute for Astrophysics and Space Research,
  Massachusetts Institute of Technology, 77 Massachusetts Avenue,
  Cambridge, MA 02139, USA}

\begin{abstract}
We present stellar mass fractions and composite luminosity functions
(LFs) for a sample of \Ncl\ clusters from the Massive and Distant
Clusters of WISE Survey (\madcows) at a redshift range of $0.951 \leq
z \leq 1.43$. Using SED fitting of optical and deep mid-infrared
photometry, we establish the membership of objects along the
lines-of-sight to these clusters and calculate the stellar masses of
member galaxies. We find stellar mass fractions for these clusters
largely consistent with previous works, including appearing to display
a negative correlation with total cluster mass. We measure a composite
\cho\ LF down to $\mstar+2.5$ for all \Ncl\ clusters. Fitting a
Schechter function to the LF, we find a characteristic \cho\ magnitude
of $\mstar=\mstarall$ and faint-end slope of $\alpha=\alphaall$ for
the full sample at a mean redshift of $\bar{z} = 1.18$. We also divide
the clusters into high- and low-redshift bins at $\bar{z}=1.29$ and
$\bar{z}=1.06$ respectively and measure a composite LF for each
bin. We see a small, but statistically significant evolution in
$\mstar$ and $\alpha$---consistent with passive evolution---when we
study the joint fit to the two parameters, which is probing the
evolution of faint cluster galaxies at $z\sim1$. This highlights the
importance of deep IR data in studying the evolution of cluster galaxy
populations at high-redshift.
\end{abstract}

\section{Introduction}
\label{Sec: Intro}
The evolution of galaxies in clusters both influences and is
influenced by the partitioning of baryons between the stars in
galaxies and the hot gas of the intracluster medium (ICM). The stellar
mass fraction of a cluster---that is, the fraction of the total mass
in stars---\fstar, offers an {\it in situ} measurement of this
partitioning. Measuring \fstar\ and its relation to other cluster
properties can give insight into the feedback processes that drive the
cycling of baryons between states and which affect how galaxies grow
and evolve \citep[\eg][]{LMS03,Ettori+06,Conroy+07}. In addition, the
shape of the cluster luminosity function (LF) offers insight into the
mass-assembly history of the cluster. The near-infrared (NIR) LF is a
useful proxy for the stellar mass function, as the luminosity in those
bands is tightly correlated with stellar mass. The NIR LF parameters
and how they evolve over time therefore reflect the mass-assembly
history of galaxies in the cluster \citep{KauffmannCharlot98}.

Previous studies such as \citet{GZZ13} have analyzed the trend of
\fstar\ with total cluster mass at $z\sim0.1$ and found an
anti-correlation. This suggests that in the local universe, larger
clusters retain their gas better and are less efficient at forming
stars. In \citet{Decker+19}, we also studied the trend of \fstar\ with
cluster mass for a sample of infrared-selected clusters at high
redshift and compared this to a sample of ICM-selected clusters at
comparable redshifts to look for differences due to selection. While
we found a larger scatter in \fstar\ in the ICM-selected clusters,
there was no significant offset between the two samples.  We also
measured a relationship between \fstar\ and total mass that was
consistent with that found by \citet{GZZ13}, but the scatter and
systematic uncertainties were too high to draw firm conclusions.

Cluster LFs follow a Schechter distribution \citep{Schechter76} of the
form
$$\Phi(m) =
\frac{\ln(10)}{2.5}\Phi^*10^{-0.4(m-\mstar)(\alpha+1)}\mathrm{exp}(-10^{-0.4(m-\mstar)}) $$
where the overall scaling is paramaterized as $\Phi^*$, the
characteristic magnitude `knee' at the bright end is parameterized as
$\mstar$ and the slope of the faint-end of the function is
parameterized as $\alpha$. Many studies have examined the rest-frame
NIR LF of galaxy clusters
\citep[\eg][]{1999AJ....118..719D,Strazzullo+06,Muzzin+07,Mancone+12,Wylezalek+14,Chan+19}
to measure the characteristic magnitude of the cluster LF at different
redshifts. However, measuring $\mstar$ becomes more difficult at
high-redshift because there is a strong degeneracy between $\mstar$
and $\alpha$.  Therefore meaningfully measuring the former at
high-redshift requires increasingly deep mid-infrared data to also
constrain the latter. Indeed, few previous studies have measured the
NIR LF for all cluster members (\ie\ those both on and off the
red-sequence) at $z>1$ down to a depth sufficient to jointly fit both
$\mstar$ and $\alpha$.

To address these problems, we measure stellar mass fractions and LFs
for a sample of clusters from the Massive and Distant Clusters of {\it
WISE} Survey \citep[\madcows,][]{Gonzalez+19}. For this work we use
\madcows\ clusters with previously measured Sunyaev-Zel'dovich
\citep[SZ,][]{SZ70,SZ72} masses, which allows us to measure \fstar\
and compare to the total mass. We also limit our sample to clusters
that have deep mid-infrared photometry. This allows us to determine
the stellar mass more robustly than in our previous study,
\citet{Decker+19}, and also allows us to measure the rest-frame NIR LF
down to sufficiently faint magnitudes to fit $\mstar$ and $\alpha$
jointly. Finally, we limit our sample to clusters that---in addition
to the above criteria---also have optical follow-up data. This allows
us to better determine which objects are true members of the clusters,
reducing systematic errors in our measurements both of \fstar\ and the
LF parameters.

We present our cluster sample and describe in more detail the
follow-up data in \textsection{\ref{Sec: sample}} and describe our
analysis in \textsection{\ref{Sec: analysis}}. Our results for both
\fstar\ and the LFs are in \textsection{\ref{Sec: results}} and we
discuss those results in \textsection{\ref{Sec:
discussion}}. Throughout this paper we use AB magnitudes in all bands
and a concordance $\Lambda$CDM cosmology of $\Omega_m = 0.3$,
$\Omega_\Lambda = 0.7$ and $H_0 = 70~\mathrm{km~s^{-1}~Mpc^{-1}}$. We
define \rfh\ as the radius inside which the cluster density is 500
times the critical density of the universe at the cluster redshift and
\mfh\ as the mass interior to that radius.

\section{Cluster Sample and Data}
\label{Sec: sample}
For this work, we use \Ncl\ clusters from the \madcows\ catalog. These
clusters are drawn from the much larger sample with SZ masses from
\citet{Brodwin+15}, \citet{Gonzalez+15}, \citet{Decker+19},
\citet{DiMascolo+20}, \citet{Dicker+20}, and
\citet{Orlowski-Scherer+21} and the SZ masses from different
facilities are generally in good agreement with each other. This
sample of clusters are selected to have previously reported
spectroscopic redshifts, deep follow-up imaging in the mid-infrared,
and optical follow-up photometry. This arrangement of follow-up data
was chosen as it allows us to constrain the membership of clusters
using photometric redshift fitting. The clusters are listed in Table
\ref{Table: cluster summary}, along with their redshifts and
information about the relevant observations. Details of the follow-up
data are given below and details of the SZ observations and mass
calculations can be found in the relevant papers.

\subsection{Optical Data}
\label{Ssec: Optical}
All \Ncl\ clusters have $r$- and $z$-band imaging from the
Gemini Multi-Object Spectrograph \citep[GMOS,][]{GMOS} on the Gemini
Telescopes in Hawai'i and Chile. These images were taken in several
programs: GN-2013A-Q-44, GN-2013B-Q-8 (both PI: Brodwin),
GN-2015A-Q-42 (PI: Perlmutter), GN-2015A-Q-4 (PI: Stalder),
GN-2017B-LP-15, GN-2018A-LP-15 (both PI: Stanford), and
GS-2019A-FT-205 (PI: Decker). There was a heterogenous mix of observing
strategies for these programs, partly due to the
different sensitivities of the GMOS CCD during different observing
cycles. However they result in a comparable depth for all the
clusters. All of the exposure times are listed in Table \ref{Table:
cluster summary}.

\subsection{Infrared Data}
\label{Ssec: Spitzer}
These clusters have mid-infrared data from the \cho\ and \cht\
bands of the {\it Spitzer Space Telescope} Infrared Array Camera
\citep[IRAC,][]{IRAC}. They were imaged in programs 12101
and 13214 (both PI: Brodwin) and the exposure times for each cluster
and each band are listed in Table \ref{Table: cluster summary}. Both
programs had the same observing strategy, with the varying exposure
times designed to detect galaxies to a relatively uniform depth
relative to $\mstar$ in different IR background regions.

\begin{deluxetable*}{lcclccccl}
  \tabletypesize{\normalsize} \tablecaption{Summary of observations of
    \madcows\ clusters used in this analysis\label{Table:
      cluster summary}} \tablewidth{0pt} \tablehead{
    \colhead{Cluster ID} & \colhead{RA} & \colhead{Dec.} &
    \colhead{$z$} & \colhead{$t_{\rm exp}$ (s)} & \colhead{$t_{\rm exp}$ (s)} &
    \colhead{$t_{\rm exp}$ (s)} & \colhead{$t_{\rm exp}$ (s)} &
    \colhead{SZ Facility} \\
    \colhead{} & \colhead{(J2000)} & \colhead{(J2000)} &
    \colhead{} & \colhead{$r$-band} & \colhead{$z$-band} &
    \colhead{\cho} & \colhead{\cht} & \colhead{} }
 \startdata
  MOO\ J0105$+$1323 & 01:05:31.5 & $+$13:23:55 & 1.143 & $13\times120$ & $9\times150$ & $14\times100$ & $33\times100$ & CARMA\tablenotemark{a} \\
  MOO\ J0319$-$0025 & 03:19:24.4 & $-$00:25:21 & 1.194 & $5\times180$ & $12\times80$ & $10\times100$ & $24\times100$ & CARMA\tablenotemark{b} \\
  MOO\ J0917$-$0700 & 09:17:04.7 & $-$07:00:08 & 1.10 & $6\times180$ & $6\times60$ & $10\times100$ & $24\times100$ & ALMA\tablenotemark{c} \\
  MOO\ J1111$+$1503 & 11:11:42.6 & $+$15:03:44 & 1.32 & $4\times300$ & $16\times80$ & $14\times100$ & $33\times100$ & CARMA\tablenotemark{a} \\
  MOO\ J1139$-$1706 & 11:39:28.2 & $-$17:06:31 & 1.31 & $12\times120$ & $10\times60$ & $14\times100$ & $34\times100$ & ALMA\tablenotemark{c} \\
  MOO\ J1142$+$1527 & 11:42:45.1 & $+$15:27:05 & 1.189 & $5\times180$ & $12\times80$ & $14\times100$ & $33\times100$ & CARMA\tablenotemark{d} \\
  MOO\ J1155$+$3901 & 11:55:45.6 & $+$39:01:15 & 1.009 & $5\times180$ & $12\times80$ & $12\times100$ & $26\times100$ & CARMA\tablenotemark{b} \\
  MOO\ J1329$+$5647 & 13:29:50.7 & $+$56:48:03 & 1.43 & $6\times180$ & $6\times60$ & $10\times100$ & $24\times100$ & GBT\tablenotemark{e} \\
  MOO\ J1506$+$5136 & 15:06:22.7 & $+$51:36:45 & 1.09 & $35\times120$ & $8\times60$ & $10\times100$ & $24\times100$ & GBT\tablenotemark{e} \\
  MOO\ J1514$+$1346 & 15:14:42.7 & $+$13:46:31 & 1.059 & $5\times180$ & $12\times80$ & $10\times100$ & $24\times100$ & CARMA\tablenotemark{b} \\
  MOO\ J1521$+$0452 & 15:21:04.6 & $+$04:52:08 & 1.312 & $12\times120$ & $9\times150$ & $12\times100$ & $26\times100$ & CARMA\tablenotemark{a} \\
  MOO\ J2206$+$0906 & 22:06:28.6 & $+$09:06:32 & 0.951 & $5\times180$ & $12\times80$ & $10\times100$ & $24\times100$ & CARMA\tablenotemark{a}
  \enddata
\tablenotetext{a}{\citet{Decker+19}}
\tablenotetext{b}{\citet{Brodwin+15}, with mass recalculated in
  \citet{Decker+19}}
\tablenotetext{c}{\citet{DiMascolo+20}}
\tablenotetext{d}{\citet{Gonzalez+15}, with mass recalculated in
  \citet{Decker+19}}
\tablenotetext{e}{\citet{Dicker+20}}
\end{deluxetable*}

%\newpage

\subsection{Catalogs}
\label{Ssec: catalogs}
For each cluster, we used the optical and infrared imaging described
in \textsection{\ref{Ssec: Optical}} and \textsection{\ref{Ssec:
Spitzer}} to make four-band ($r$, $z$, \cho, \cht) photometric
catalogs. For each cluster, all four images were transformed onto the
same image scale using SWarp \citep{SWarp}. The catalogs were produced
by running Source Extractor \citep[SE,][]{SExtractor} in dual-image
mode on all four SWarped images, using the \cho\ image as the
detection image. The SE parameters were the same as in
\citet{Decker+19}. The final catalogs used $2\arcsec$ diameter
aperture photometry in the optical bands and $4\arcsec$ diameter
corrected to $24\arcsec$ diameter aperture photometry in the infrared
bands. The correction from $4\arcsec$ to $24\arcsec$ used the IRAC
aperture corrections given in \citet{Ashby+09}.

For comparison and validation of our fitting (see next section) we
also made a field catalog using $r$- and $z$- band images from the
Cosmic Evolution Survey \citep[COSMOS,][]{COSMOS} and \cho\ and \cht\
images from the \spitzer\ Extended Deep Survey \citep{SEDS} in the
footprint where those two surveys overlap. These catalogs were made
with the same procedure as our cluster catalogs. Because the \spitzer\
Extended Deep Survey imaging is deeper than our IRAC imaging, we
artificially degraded the field catalog data in the IRAC bands to
match our cluster catalogs. We do this by adding a small additional
scatter to the measured fluxes. This scatter is randomly drawn from a
Gaussian distribution with a width equal to the quadrature difference
of the (higher) error in our photometry and the error in the Extended
Deep Survey.

%\newpage

\section{Analysis}
\label{Sec: analysis}

\subsection{Cluster Membership}
\label{Ssec: membership}
We used EAZY \citep{EAZY} to fit spectral energy distributions (SEDs)
to the four-band photometry for each object in our cluster catalogs
and our field catalog. The result was a best-value and a probability
density function (PDF) of the redshift for each object. We compared
the best-value fitted redshifts in the field catalog to the multi-band
photometric redshifts from the COSMOS catalog \citep{Laigle+16} to
determine the error in our photometric redshift fitting. After running
an iterative $3\sigma$ clipping routine, we found an error in our
photometric redshifts of $\sigma_z = 0.17(1+z)$. This error is
relatively high, due to the small number of photometric bands, but is
sufficient to isolate cluster members with a low interloper rate.

\begin{figure}[bthp]
\center{\includegraphics[width=8.5cm]{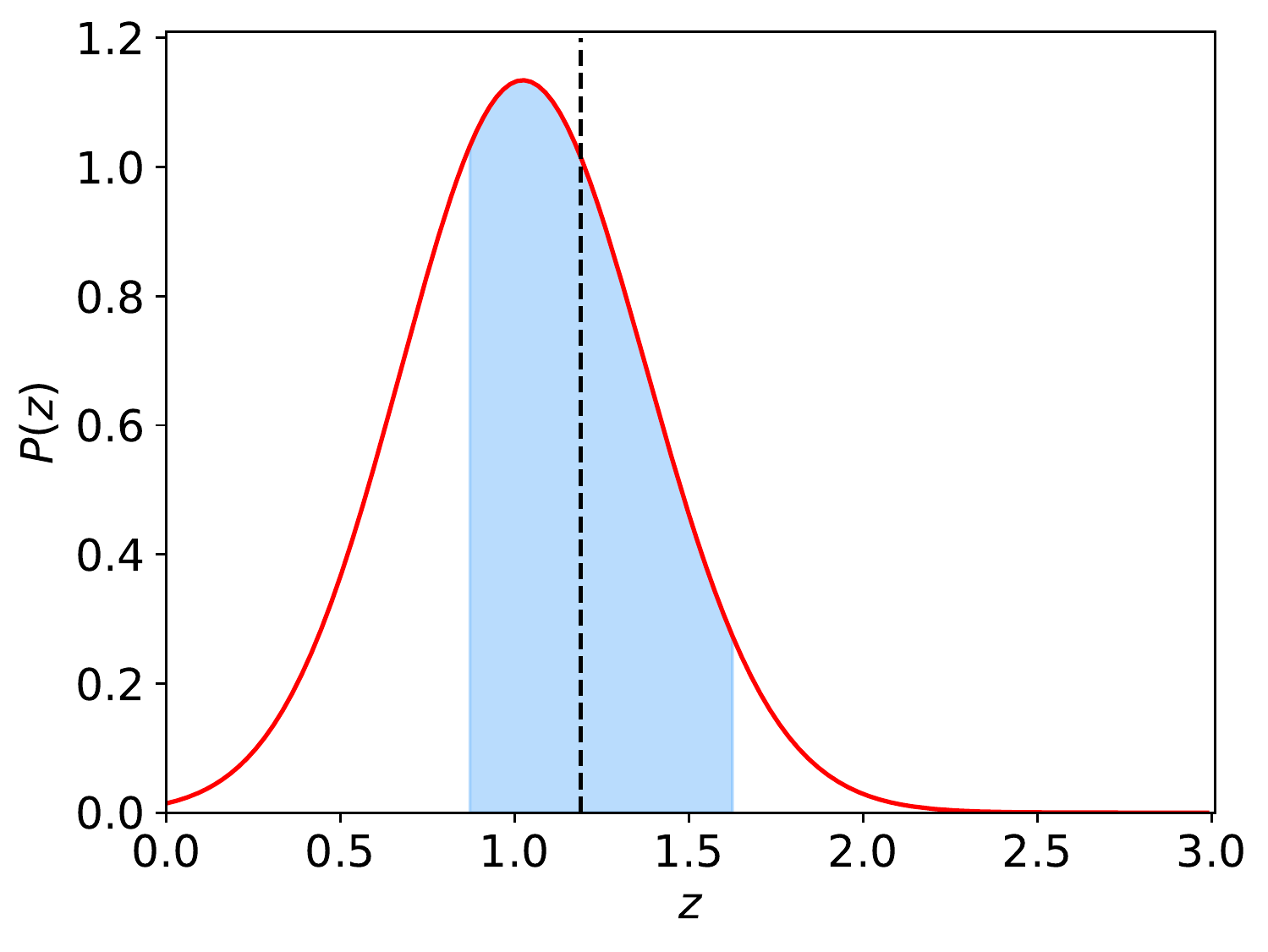}}
\caption{Smoothed output PDF from EAZY of a spectroscopically-confirmed
  \citep[from][]{Gonzalez+15} member of MOO\ J1142$+$1527. The cluster
  redshift of $z=1.19$ is denoted by the black dashed line and the
  shaded blue region is the integration range around the cluster
  redshift. The integrated probability in that region is 0.65, well
  above our membership threshold.}
\label{Fig: PDF integration}
\end{figure}

To determine which objects in the cluster catalogs were consistent
with being members of the cluster, we used the full PDFs output by
EAZY. For each object, we first smoothed the output PDF with a
Gaussian corresponding to the $\sigma_z=0.17(1+z)$ scatter in our
redshift fitting. We then integrated under this convolved PDF in the range
$z_{cl} - \sigma_{z} \leq z < z_{cl} + \sigma_{z}$, as shown in Figure
\ref{Fig: PDF integration}. If this integrated probability was above
0.3, we considered the object to be a cluster member; everything else
was removed from the catalog. We chose 0.3 as a cutoff to maximize
completeness while still removing the bulk of the line-of-sight
interlopers. Because our cluster masses were only measured out to
\rfh, we also removed from our cluster catalogs objects lying at a
projected distance more than that distance from the cluster
center. Since this method would still not remove every line-of-sight
interloper, we also ran this analysis on the field catalog at each
cluster redshift. This provided a set of \Ncl\ field catalogs, each
containing the objects from the full field catalog that our analysis
would consider being consistent with cluster members. These
`interloper' catalogs provided us a baseline that allowed us to
statistically remove line-of-sight interlopers from our analysis.

\subsection{Completeness}
\label{Ssec: completeness}
\begin{figure}[bthp]
\center{\includegraphics[width=8.5cm]{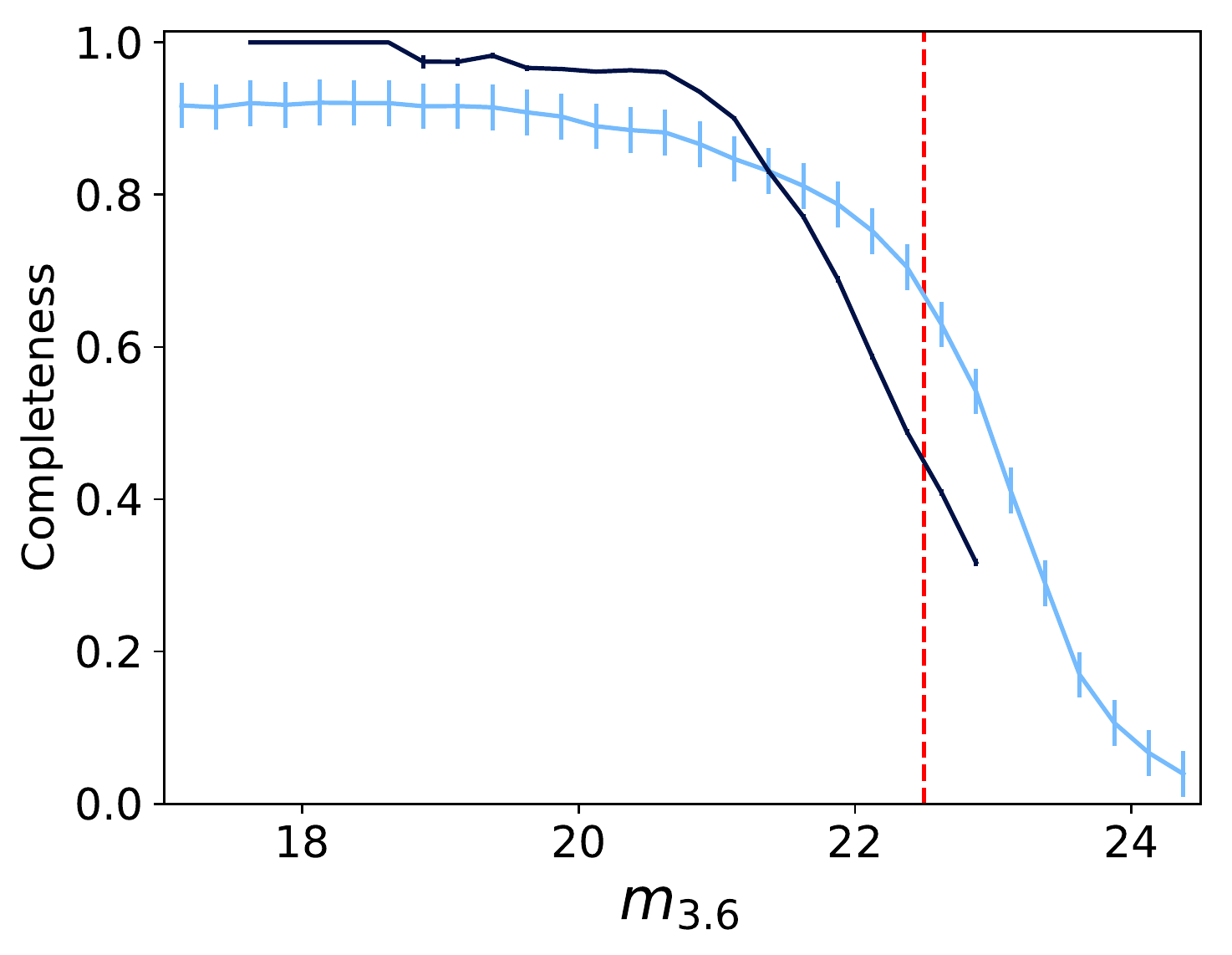}}
\caption{Mean detection (light blue) and membership (dark blue)
completeness for the clusters in our sample. The error bars on the
detection completeness are Poisson errors and the error bars on the
membership completeness are from bootstrap resampling. At the bright
end of the membership completeness curve (\cho$< 18.75$) there were not
enough field objects to produce meaningful statistics, so we fixed the
completeness to unity. The vertical dashed red line at the faint end
represents the average $5\sigma$ detection limit.}
\label{Fig: all completeness}
\end{figure}

We measured the photometric completeness of our cluster catalogs by
randomly placing artificial sources into our detection images, running
SE, and recording how many of these artificial sources were detected
by SE. For each cluster we placed a total of one thousand sources per
quarter-magnitude bin in batches of ten sources each. The average
\cho\ completeness curve of the sample is shown as a light blue line
in Figure \ref{Fig: all completeness}. The completeness reaches a
plateau at around 95\% at the bright end because of the high density
of infrared sources in the clusters. The vertical dashed line
represents the average $5\sigma$ limit of our data in that
cluster. The errors on the completeness in each bin are Poisson errors
and are approximately $3\%$ per bin.

In addition to measuring the photometric completeness of our catalogs,
we also measured the completeness of our cluster member selection
algorithm. Each object in the field catalogs has a redshift from
COSMOS, and some coincidentally lie at the redshifts of our
clusters. For each cluster, we isolated these objects from the field
catalog and ran our membership selection algorithm on them. As with
the detection completeness, we split the objects into
quarter-magnitude bins in \cho. For each quarter-magnitude bin, we
define the membership completeness as the fraction of objects that our
algorithm correctly identified as lying at the cluster redshift. Since
unlike with the artificial sources we used for our detection
completeness, there were a variable number of objects in each bin, we
calculated the error on the membership completeness using bootstrap
resampling. The average membership completeness is shown in Figure
\ref{Fig: all completeness} as a dark blue line. Since there were not
enough objects at the bright end to have meaningful statistics, we
fixed the completeness in that region to unity.

\subsection{Stellar Mass}
\label{Ssec: mass fitting}
We used another SED-fitting program, FAST \citep{FAST}, on the cluster
catalogs to calculate the stellar masses of the objects along the line
of sight to the cluster. For this, we adopted a \citet{BC03} model, a
\citet{Chabrier03} initial mass function, a solar metallicity, and we
fixed the redshift of each object in the catalog to the cluster
redshift. We only fit to intrinsic properties of the galaxies, in
particular stellar mass. As with the redshift fitting, we also ran
FAST on the field catalog to calibrate the errors in our fitting and
to establish how much field contribution to expect even after removing
interlopers. Comparing the stellar masses we measured in this way to
the stellar masses given in the COSMOS catalog, we adopted a uniform
$0.2~\mathrm{dex}$ uncertainty in our stellar mass measurements.

%If it comes up, FAST was fitting a tau model, with log tau between
%8.5 and 10, and fitting the age of the object, with log age between
%8.0 and 10.

With this final catalog of objects identified as being at the cluster
redshift by EAZY, lying within a projected distance of \rfh\ from the
cluster center, and with stellar masses measured from FAST, we
calculated the total stellar mass of the clusters inside \rfh. For
each cluster, we first scaled the stellar mass of each object by the
photometric and membership completeness corrections in
\textsection{\ref{Ssec: completeness}}. We then summed these scaled
masses to get a total line-of-sight stellar mass for the
cluster. Since this is still expected to include some small number of
interlopers, we measured the total stellar mass of the statistical
interloper catalog in the same way. We scaled that mass to the area of
the cluster catalog and subtracted this expected interloper
contribution---approximately 10\% of the line-of-sight mass for most
of the clusters---from the line-of-sight mass to get the total cluster
stellar mass. We calculated the error on the stellar mass of each
cluster by propagating the error of the stellar mass of each
individual object in the cluster and field catalogs, measured in
\textsection{\ref{Ssec: membership}}.

%\newpage

\subsection{Luminosity Function}
\label{Ssec: lf}
We used our membership selection and deep IRAC photometry to produce
composite \cho\ LFs for our cluster sample. For each cluster, we first
evolution-corrected the \cho\ apparent magnitudes of both the cluster
and interloper catalog to the mean redshift of the sample using EZGal
\citep{EZGal} and assuming passive evolution after an initial
starburst at $z_f = 3.0$. We removed the brightest cluster galaxy
(BCG) from the cluster catalog and then binned these
evolution-corrected catalogs into quarter-magnitude bins to produce a
line-of-sight LF and a background LF for each cluster. We then applied
both the completeness corrections described in \textsection{\ref{Ssec:
completeness}} as a function of magnitude to both LFs. Finally, we
scaled the background LF to match the surface area of the cluster and
subtracted it off the line-of-sight LF to produce the individual
cluster LF. These individual LFs were stacked to form the composite LF
for the sample. The error on each value in the individual LFs is from
adding in quadrature the Poisson errors of both the line-of-sight and
interloper LFs and the errors on both completeness corrections. The
error on each value in the composite LF is the quadrature sum of those
errors from the individual LFs.

\section{Results} 
\label{Sec: results}
\subsection{Stellar Mass Fractions}
\label{Ssec: fstar}
\begin{figure*}[bthp]
\center{\includegraphics[width=18cm]{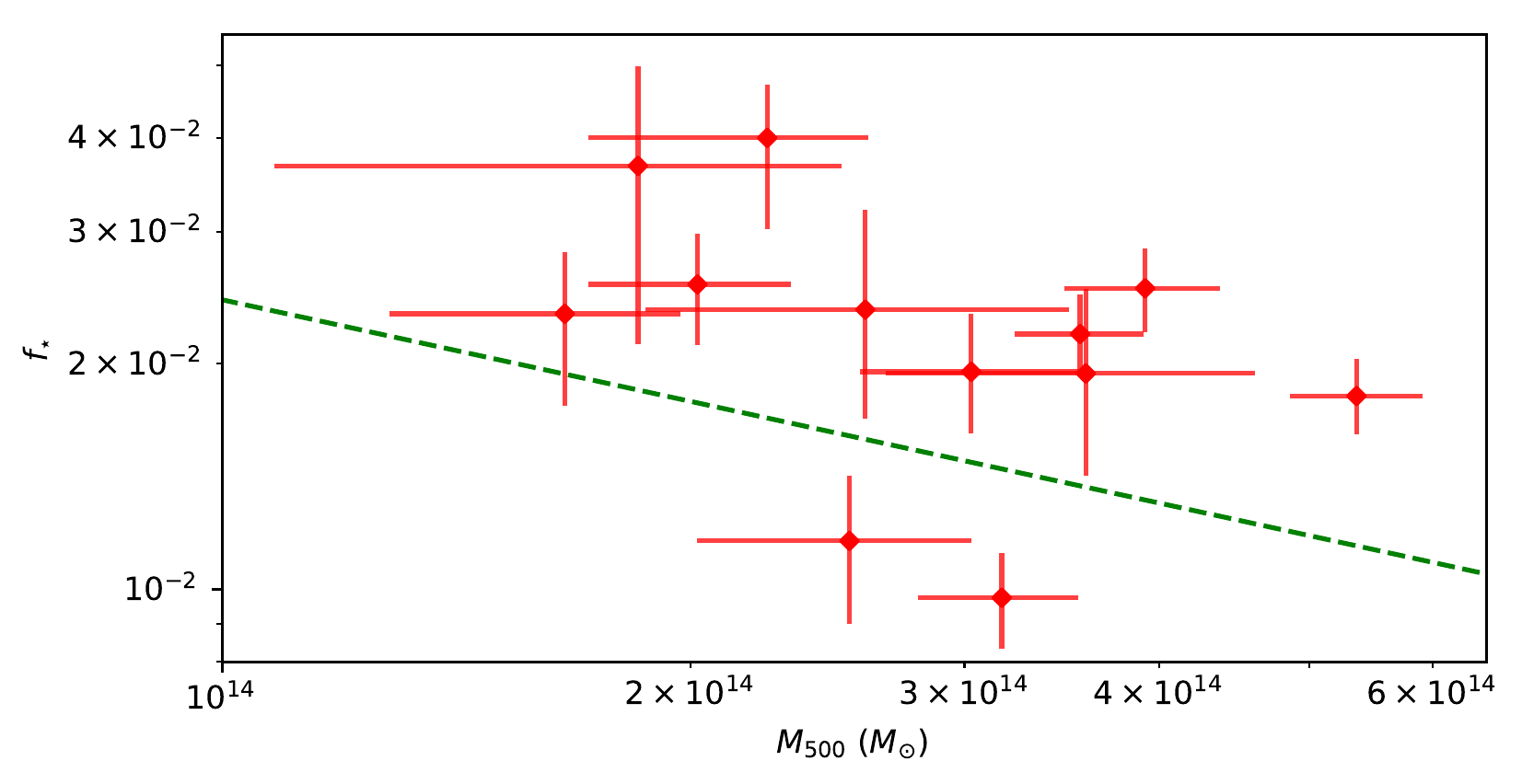}}
\caption{Stellar mass fraction versus total mass for the
\madcows\ clusters in this analysis, plotted as red diamonds. The error
on \fstar\ is from adding the percent stellar mass error from
\textsection{\ref{Ssec: mass fitting}} and the percent error of the
total mass of the cluster in quadrature. The green dashed line is the
low-redshift relation from \citet{GZZ13}, plotted to provide continuity
with \citet{Decker+19}.}
\label{Fig: fstarvm500}
\end{figure*}

The stellar mass fractions we measure for these clusters are given in
Table \ref{Table: fstars} and Figure \ref{Fig: fstarvm500} shows
\fstar\ versus \mfh\ for the \Ncl\ clusters of this work, plotted as
red diamonds. For comparison, we also plot the low-redshift trend line
measured by \citet{GZZ13} as a green dashed line.  Clusters that were
also studied in \citet{Decker+19} are indicated in Table \ref{Table:
fstars}.

\begin{deluxetable}{lccc}
  \tabletypesize{\normalsize} \tablecaption{\madcows\ Stellar Mass Fractions \label{Table: fstars}} \tablewidth{0pt} \tablehead{
    \colhead{ID} & \colhead{\mfh} &
    \colhead{$M_\star$} & \colhead{\fstar}\\
    \colhead{} & \colhead{$(10^{14}~\msol)$} &
    \colhead{$(10^{12}~\msol)$} & \colhead{$(10^{-2})$}}
  \startdata
  MOO\ J0105$+$1323* & $3.92^{+0.46}_{-0.44}$ & $10.6\pm0.58$ & $2.70^{+0.35}_{-0.34}$ \\
  MOO\ J0319$-$0025* & $3.03^{+0.54}_{-0.46}$ & $6.32\pm0.49$ & $2.09^{+0.40}_{-0.36}$ \\
  MOO\ J0917$-$0700 & $1.66^{+0.31}_{-0.38}$ & $4.08\pm0.36$ & $2.48^{+0.51}_{-0.61}$ \\
  MOO\ J1111$+$1503* & $2.02^{+0.30}_{-0.30}$ & $4.95\pm0.40$ & $2.45^{+0.42}_{-0.42}$ \\
  MOO\ J1139$-$1706 & $2.24^{+0.36}_{-0.52}$ & $8.40\pm0.66$ & $3.80^{+0.68}_{-0.93}$ \\
  MOO\ J1142$+$1527* & $5.36^{+0.55}_{-0.50}$ & $9.91\pm0.59$ & $1.85^{+0.22}_{-0.20}$ \\
  MOO\ J1155$+$3901* & $2.53^{+0.50}_{-0.51}$ & $3.21\pm0.29$ & $1.27^{+0.28}_{-0.28}$ \\
  MOO\ J1329$+$5647 & $3.56^{+0.35}_{-0.33}$ & $7.89\pm0.64$ & $2.25^{+0.28}_{-0.27}$ \\
  MOO\ J1506$+$5136 & $3.17^{+0.38}_{-0.37}$ & $3.59\pm0.27$ & $1.16^{+0.16}_{-0.16}$ \\
  MOO\ J1514$+$1346* & $2.39^{+0.51}_{-0.83}$ & $7.34\pm0.49$ & $3.97^{+1.42}_{-1.67}$ \\
  MOO\ J1521$+$0452* & $3.59^{+1.02}_{-0.92}$ & $7.15\pm0.60$ & $1.99^{+0.59}_{-0.54}$ \\
  MOO\ J2206$+$0906* & $2.95^{+0.82}_{-0.68}$ & $6.52\pm0.40$ & $2.52^{+0.90}_{-0.72}$
  \enddata
\tablenotetext{*}{Also in \citet{Decker+19}}
\end{deluxetable}

\begin{figure}[bthp]
\center{\includegraphics[width=8.5cm]{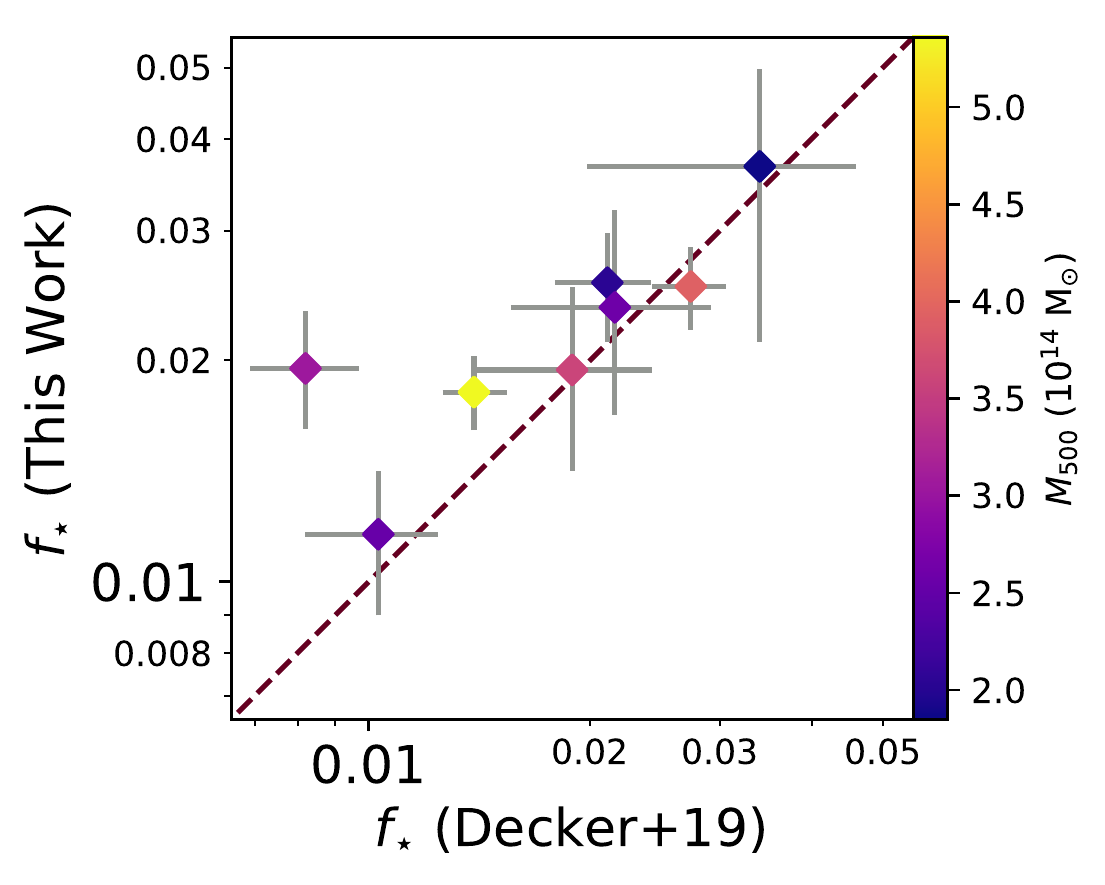}}
\caption{Comparison of the stellar mass fractions of clusters common
to this work and to \citet{Decker+19}. The dashed red line represents
where the old and new mass fractions would be equal. Clusters falling
above the line have a higher \fstar\ in this work and clusters falling
below the line have a higher \fstar\ in \citet{Decker+19}. The color
of each point corresponds to the mass of the cluster.}
\label{Fig: oldcomp}
\end{figure}

Figure \ref{Fig: oldcomp} shows the direct comparison of \fstar\ for
the eight clusters common to both this work and \citet{Decker+19}.
With only one exception, MOO\ J0105$+$1323, the \fstar\ we measure in
this work is higher than the \fstar\ we found in \citet{Decker+19} and
for no clusters is it significantly lower. This is expected as in both
works we use IRAC \cho\ to measure \fstar---either directly or
indirectly---and the much deeper \cho\ data in this work allow us to
include stellar mass from galaxies that were too faint to be Using
\cho\ luminosity as a proxy for stellar mass, we quantify the amount
of `extra' stellar mass we should expect with these deeper data by
integrating down the composite LF we measure in
\textsection{\ref{Ssec: LF results}}. Integrating down to the depth of
our current data versus integrating to the depth of our data in
\citet{Decker+19} shows we are sensitive to approximately 25\% more
stellar mass with these deep IRAC data than we were previously. This
is consistent with the change in \fstar\ we see in all but two of the
clusters. Further integrating the LF arbitrarily deep shows this
analysis is sensitive to $\geq 95\%$ of the stellar mass in each
cluster.

MOO\ J0319$-$0025 and MOO\ J1142$+$1527 exhibit larger jumps in
\fstar\ than the 25\% we expect simply from the deeper data. Those
increases likely come from the improved way we are determining both
cluster membership and stellar mass.  Using a fuller sampling of the
galaxy SEDs---even in just four bands---gives us a better and more
consistent measurement of the stellar mass of each object versus what
we were able to do in \citet{Decker+19}.

\subsection{Luminosity Functions}
\label{Ssec: LF results}

\begin{deluxetable*}{lcccccc}
  \tabletypesize{\normalsize} \tablecaption{\madcows\ Sample Data and
    Schechter Parameters \label{Table: Schechter parameters}} \tablewidth{0pt} \tablehead{
    \colhead{Sample} & \colhead{N$_{cl}$} & \colhead{$\bar{z}$} & \colhead{$\bar{M_{500}}$} &
    \colhead{$\mstar$} & \colhead{$\alpha$} & \colhead {$\phi^*$}\\
    \colhead{} & \colhead{} & \colhead{} & \colhead{($10^{14}\msol$)} &
    \colhead{} & \colhead{} & \colhead{($\mathrm{dN}~\mathrm{dm}^{-1}~\mathrm{cluster}^{-1}$)}}
  \startdata
  All & 12 & 1.18 & $2.96$ & $\mstarall$ & $\alphaall$ & $98\pm13$ \\
  High-$z$ & 6 & 1.29 & $3.30$ & $\mstarhigh$ & $\alphahigh$ & $89\pm20$ \\
  Low-$z$ & 6 & 1.06 & $2.62$ & $\mstarlow$ & $\alphalow$ & $96\pm20$
  \enddata
\end{deluxetable*}

\begin{figure*}[bthp]
\center{\includegraphics[width=18cm]{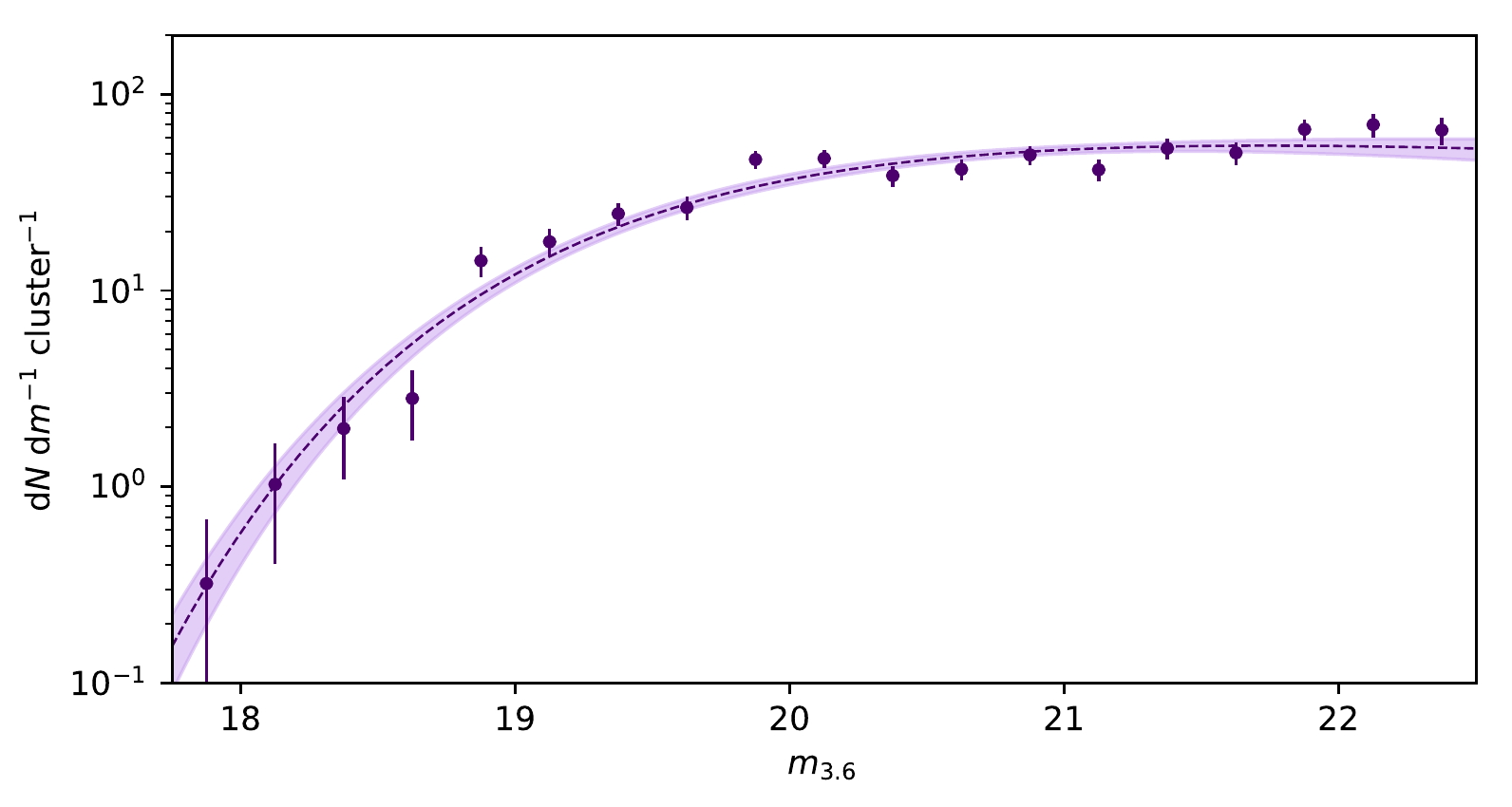}}
\caption{Composite LF for the full sample of \Ncl\ clusters used in this
analysis. The purple points are the average number of objects per
magnitude per cluster in each quarter-magnitude bin and the error bars
are derived from the Poisson error of the individual LFs and the
completeness errors. The mean sample mass and redshift are given, as
well as the best fits to the parameterized Schechter function and
their errors, in Table \ref{Table: Schechter parameters}. The best-fit
Schechter function itself is shown as a dashed line and the shaded
area represents the $1\sigma$ error on the best fit.}
\label{Fig: lf_w_fit_all}
\end{figure*}

The composite LF for our full sample of \Ncl\ clusters is shown in
Figure \ref{Fig: lf_w_fit_all}. We fit a parameterized Schechter
function to the measured LF using a Monte Carlo Markov Chain (MCMC)
running a Metropolis-Hastings algorithm. The best-fit Schechter
function is shown as a dashed line in the figure, with the lighter
region showing the $1\sigma$ error on the best fit. The best fit
parameters and the error on them are derived from the mean and
standard deviation of the MCMC posterior chains after discarding the
initial `burn-in' period. These values are listed in Table \ref{Table:
Schechter parameters} along with the mean redshift and mass of the
sample. 

%%%ORIGINAL SITE OF TABLE 3%%%

\begin{figure}[bthp]
\center{\includegraphics[width=8.5cm]{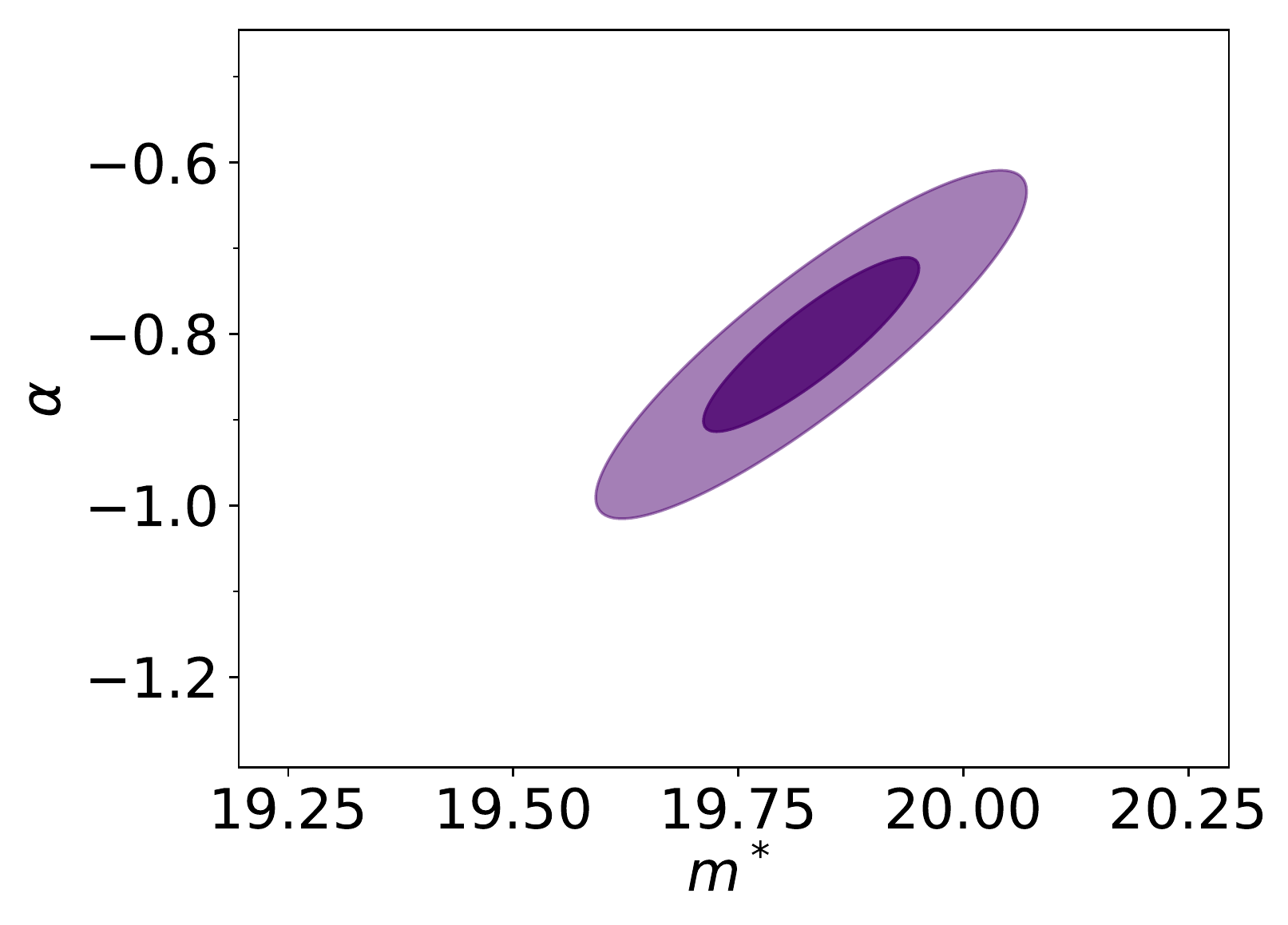}}
\caption{Covariance ellipses showing the $1\sigma$ (dark) and
  $2\sigma$ (light) errors and covariances in $\mstar$ and $\alpha$
  for the average LF of all \Ncl\ clusters in this work.}
\label{Fig: covariance_all}
\end{figure}

Because $\mstar$ and $\alpha$ are covariant, in addition to
the simple errors given in Table \ref{Table: Schechter parameters}, we
also plot the $1\sigma$ (dark) and $2\sigma$ (light) covariance
ellipses for $\mstar$ and $\alpha$ for the full sample in Figure
\ref{Fig: covariance_all}. This shows the extent of the degeneracy
between $\mstar$ and $\alpha$ as well as the axis along which our
uncertainty is concentrated.

\begin{figure*}[bthp]
\center{\includegraphics[width=18cm]{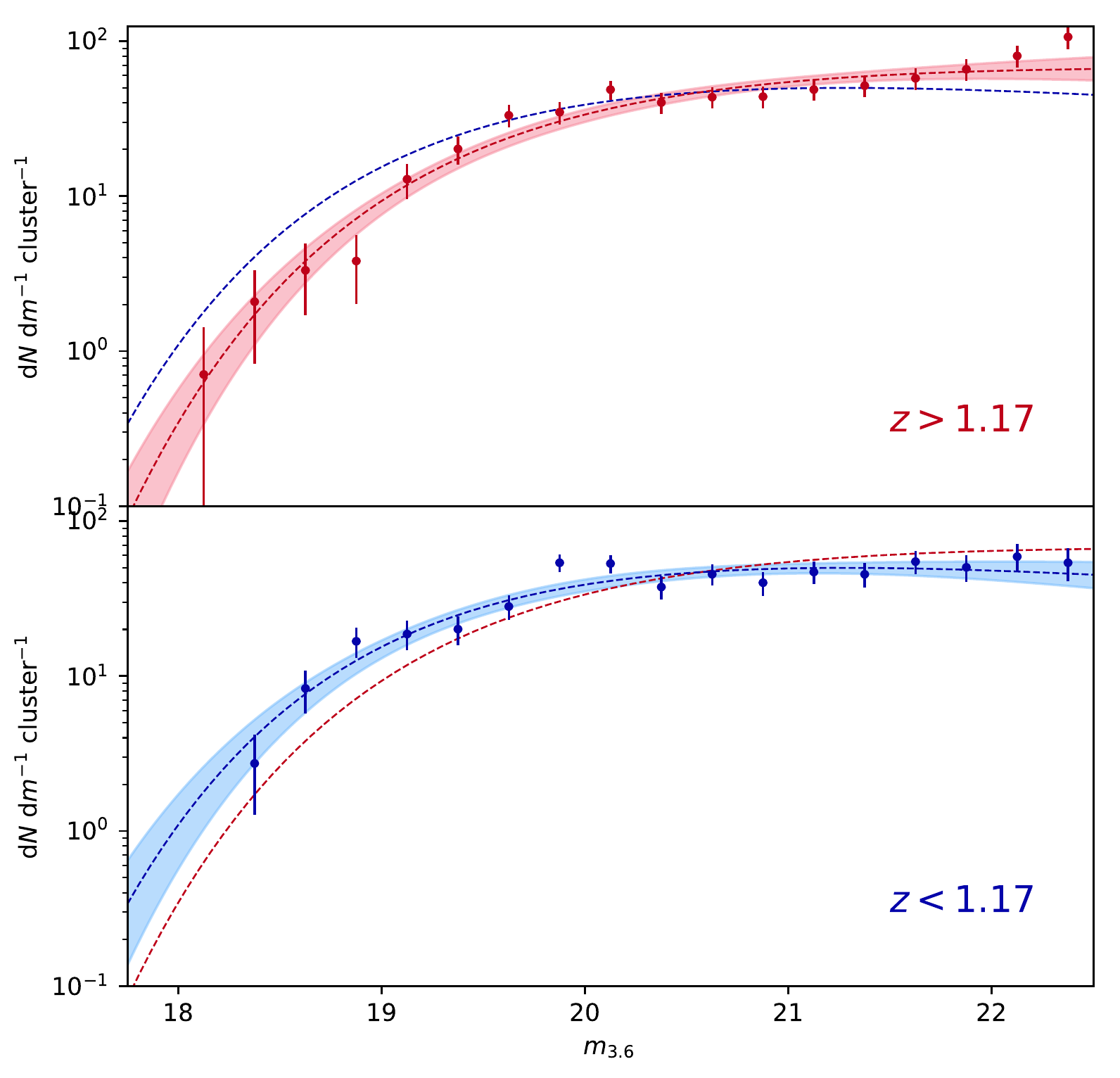}}
\caption{LFs for the high-redshift (upper, red) and low-redshift
  (lower, blue) subsamples. The best-fit Schechter functions are
  plotted as dashed lines with the $1\sigma$ error on the best fit
  shown as shaded regions. Each panel also shows the best fit line
  from the other panel as a dashed line of the relevant color.}
\label{Fig: lf_w_fit_subsamples_z}
\end{figure*}

Our sample is large enough that we also split our clusters into high-
and low-redshift samples---splitting them at the median redshift of
our sample, $z=1.17$---and measure composite LFs for both of
those. The measurement and fitting for these sub-samples is the same
as for the full sample and the mean masses and redshifts for both
sub-samples are also shown in Table \ref{Table: Schechter parameters},
along with the best-fit Schechter function parameters to each
LF. These LFs are shown in Figure \ref{Fig: lf_w_fit_subsamples_z}.
As with the full sample, we also plot the covariance between $\mstar$
and $\alpha$ for these two sub-samples in Figure \ref{Fig:
covariance_z_comp}.  This figure shows that although the error bars
for the individual parameters overlap between the two samples, there
is a significant evolution from $z=1.29$ to $z=1.06$ in the LF as a
whole. As discussed in \textsection{\ref{Sssec: LF evolution}}, this
is consistent with being driven by passive evolution in our
sample. For comparison, we also plot in Figure \ref{Fig:
covariance_z_comp} the outlines of the ellipses for the full sample
from Figure \ref{Fig: covariance_all}.

\begin{figure}[bthp]
\center{\includegraphics[width=8.5cm]{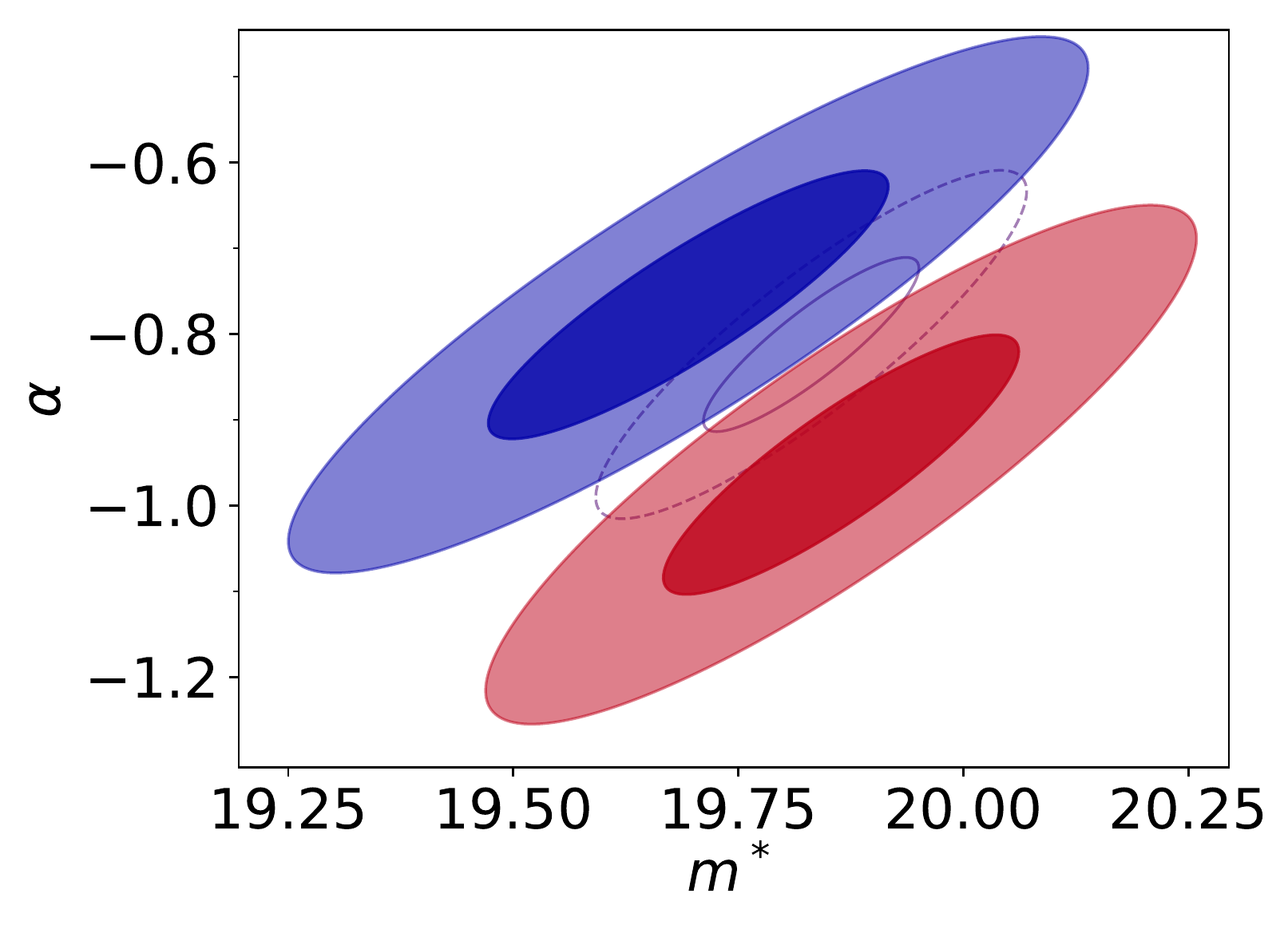}}
\caption{Covariance ellipses showing the $1\sigma$ (dark) and
$2\sigma$ (light) errors and covariances in $\mstar$ and $\alpha$ for
the high-redshift (red) and low-redshift (blue) sub-samples. For comparison, the
whole-sample covariances ellipses from Figure \ref{Fig: covariance_all}
are outlined underneath.}
\label{Fig: covariance_z_comp}
\end{figure}

We also explore fitting a sum of two Schechter function to our LFs, in
a manner similar to \citet{Lan+16}. This is motivated particularly by
our high-redshift LF which seems to show an upturn at the faint end
that is possibly more consistent with a second `faint' Schechter
function wit a steep faint-end slope. Although we can't rule out there
being an upturn at the faint-end of our LFs, to the depth of our data
($\sim \mstar+2.5$) we find that this sum of Schechter functions is
at best only a marginally better fit to the data, at a level that is
well short of statistical significance.

\section{Discussion}
\label{Sec: discussion}
\subsection{Stellar Mass Fraction}
\label{Ssec: fstar discussion}
The stellar mass fractions we compute in this work display many of the
same traits as the stellar mass fractions we calculated in
\citet{Decker+19}. The main difference is that the improved
measurement of \fstar\ has resulted in higher values overall, and ten
of the twelve now lie slightly above the \citet{GZZ13} line, nine of them
significantly so. They still appear to follow the slope of the
\citet{GZZ13} line, however. Despite our improved measurements
significantly reducing the systematic errors on our measurements of
\fstar, the statistical errors are still high enough that we cannot
draw any significant conclusions about either the slope of the trend
or its normalization. Much of this is due to the high error on the
total mass of some of the clusters, which factors into the error on
\fstar. For the well-measured clusters like MOO\ J1142$+$1527 on the
right side of Figure \ref{Fig: fstarvm500}, the \fstar\ error is very
small. Deeper SZ imaging on the \madcows\ clusters is likely necessary
to provide a significant measurement of properties related to the
total mass.

%\newpage

\subsection{Comparison to Other LF Studies}
\label{Ssec: comparison}
\begin{figure}[bthp]
\center{\includegraphics[width=8.5cm]{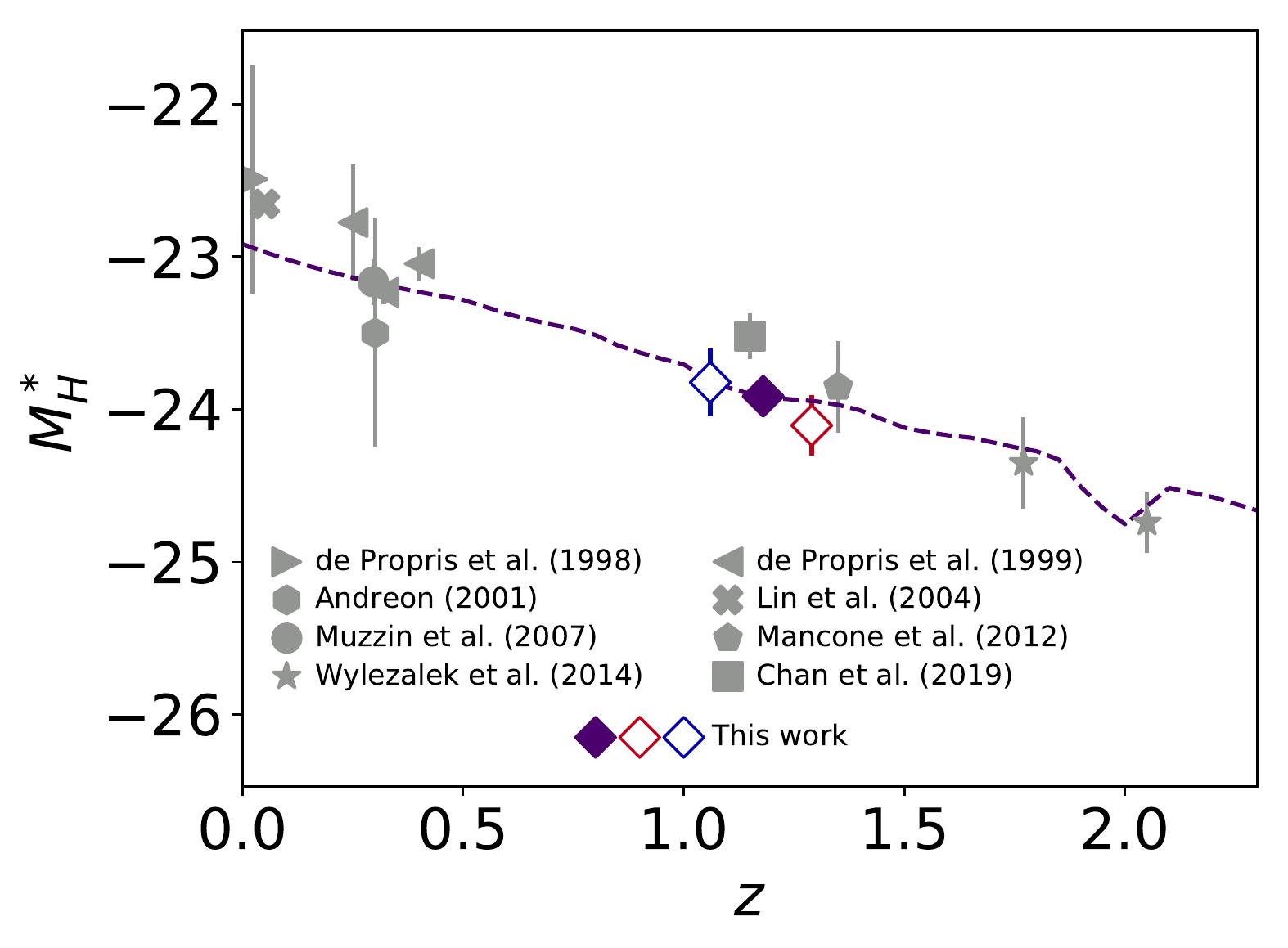}}
\caption{Absolute $H$-band characteristic magnitude versus redshift
  for a number of cluster LF studies. The error on the \citet{LMS04} point is
  0.02, smaller than the data marker. The dashed purple line shows the
  expected passive evolution in $M^*_H$}
\label{Fig: mstar v z}
\end{figure}

\begin{figure}[bthp]
\center{\includegraphics[width=8.5cm]{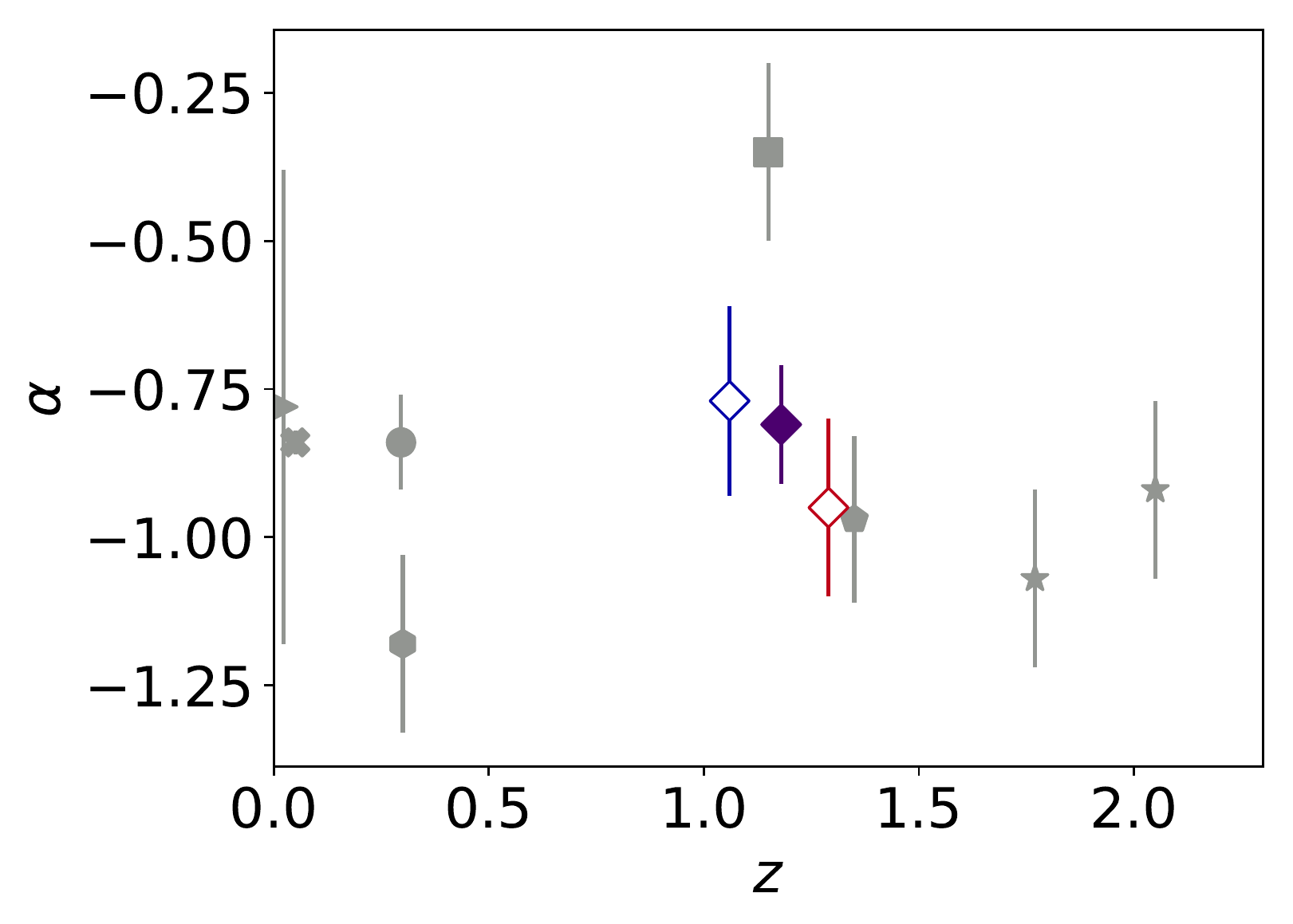}}
\caption{Faint-end slope versus redshift for a number of cluster LF
studies. The labels for the points are the same as in Figure \ref{Fig:
mstar v z}. The error on the \citet{LMS04} point is 0.02, smaller than
the data marker.}
\label{Fig: alpha v z}
\end{figure}

A comparison of our measurements of $\mstar$ and $\alpha$ to other
works across a range of redshifts is shown in Figures \ref{Fig: mstar
v z} and \ref{Fig: alpha v z}. To facilitate comparison with some
other studies, we again use EZGal to convert our apparent
\cho\ magnitudes to absolute $H$-band magnitudes. We use a
\citet{BC03} model with a formation redshift of $z_f=3.0$, a
\citet{Chabrier03} initial mass function, and a solar metallicity to
calculate the $k$-correction, but because we are already probing the
rest-frame $H$-band with our \cho\ observations, the $k$-correction is
almost entirely model-independent and changing the model parameters
does not affect the $k$-correction by more than $0.01$ at any of our
redshifts. Doing this, we find $M_H^*= \MHall$ for our overall LF and
$M_H^*= \MHhigh$ and $M_H^*= \MHlow$ for the high- and low-redshift
LFs, respectively. Figure \ref{Fig: mstar v z} shows this comparison
for $M_H^*$.

For both $M^*$ and $\alpha$ we plot the results from our full sample
as a solid diamond and the results from the high- and low-redshift
samples as red and blue open diamonds, respectively. In Figure
\ref{Fig: mstar v z}, we also plot as a dashed line the expected trend
of $M^*_H$ with redshift, assuming passive evolution. We discuss the
comparison works below.

\subsubsection{Comparisons at Similar Redshift}
\label{Sssec: direct lf comp}
Owing to the difficulty in getting deep enough mid-infrared data, few
studies have previously been able to measure both $\mstar$ and
$\alpha$ simultaneously at this redshift range. One such study is
\citet{Mancone+12}, who measured composite LFs in \cho\ and \cht\ for
seven IRAC Shallow Cluster Survey (ISCS) clusters at a median redshift
of $z=1.35$, slightly higher than our mean redshift overall, but a
good match to our high-redshift sub-sample. Converting the apparent
Vega magnitudes they report into absolute AB magnitudes, they found
best fit parameters for the \cho\ LF of $M_H^*=-23.85\pm0.30$ and
$\alpha=-0.97\pm0.14$. The faint-end slope we measure for the
high-redshift clusters matches this result almost exactly and the
characteristic magnitude we find is also consistent with their
results.

Another similar study is \citet{Chan+19}, who used IRAC \cho\ imaging
to measure rest-frame $H$-band LFs for red-sequence galaxies in seven
clusters from the infrared-selected Gemini Observations of Galaxies in
Rich Early Environments \citep[GOGREEN,][]{GOGREEN} survey at a mean
redshift of $\bar{z}=1.15$. They report their results in terms of
absolute $H$-band magnitudes and find $M_H^*=-23.52^{+0.15}_{-0.17}$,
which is somewhat fainter than our value of $M_H^*= \MHall$ at
$z=1.18$. They also find a much more steeply falling faint-end slope
than we do, with a value of $\alpha=-0.35\pm0.15$.  This difference in
$\alpha$ may be a result of their only including red-sequence
galaxies, whereas we include everything with a photometric redshift
consistent with being a cluster member. Other works
\citep[\eg][]{Muzzin+07,Strazzullo+06,dePropris17} have found that the
faint-end of the cluster LF is dominated by blue galaxies, which would
explain the discrepancy between our results and red-sequence-only
results. Alternatively, \citet{Connor+19} found that the traditional
model of the red sequence as a sloped line does not hold at fainter
magnitudes. This could cause a drop-off in the measured red-sequence
LF that is unconnected to the galaxy population of the cluster.

%\newpage

\subsubsection{Comparisons at Other Redshifts}
\label{Sssec: low-z lf comp}
To put our results into a wider context, we also compare to studies at
other redshifts and with different cluster selection and fitting methods. To
make as fair a comparison as possible to $\mstar$ at other redshifts, we use
the model described in \textsection{\ref{Ssec: comparison}} to
convert all the magnitudes into the absolute $H$-band and limit
ourselves to studies where the reported values still probe either the
rest-frame $H$-band or nearby $K$-band. These comparisons are also
shown in Figures \ref{Fig: mstar v z} and \ref{Fig: alpha v z}.

At the lowest-redshift of our comparisons, \citet{dePropris+98} looked
at the Coma cluster at $z=0.023$ to a depth much fainter than
$\mstar$. They jointly fit $\mstar$ and $\alpha$ at magnitudes
brighter than $\mstar+3$, though at magnitudes fainter than
$\mstar+3$, they found a sharp rise in the number of galaxies and fit
this with a power law. At a similar redshift, \citet{LMS04} calculated
a stacked $K$-band LF for a sample of 13 Abell clusters with X-ray
follow-up. We compare to their joint fits of $\mstar$ and $\alpha$,
though they also attempt to fix $\alpha$ due to the uncertainty in the
faint-end slope. Their errors on both parameters are $0.02$, which is
small enough it does not appear on Figures \ref{Fig: mstar v z} and
\ref{Fig: alpha v z}. At higher redshift, \citet{1999AJ....118..719D}
looked at a heterogenous selection of clusters in the $K$-band in
redshift bins up to $z=0.92$. They only fit to $\mstar$, fixing the
faint-end slope at $\alpha=-0.9$. We compare $z=0.25$, $z=0.32$ and
$z=0.4$ bins, where the conversion from $K$-band into rest-frame
$H$-band has a minimal $k$-correction. Similarly, \citet{Muzzin+07},
measured the observed-frame $K$-band for clusters at a mean redshift
of $\bar{z}=0.296$. In addition to reporting a composite LF for all
galaxies in their clusters, \citet{Muzzin+07} also split their
galaxies by whether they were quiescent or
star-forming/recently-quenched. They found a relatively flat faint-end
slope for their overall LF, $\alpha=-0.84\pm0.08$ and a much steeper
faint-end slope of $\alpha=0.17\pm0.18$ for red-sequence galaxies.  At
similar redshift, \citet{Andreon01} studied a single cluster at
$z=0.3$ in the $K_S$-band. He measured a LF down to $\mstar+5$ in
various areas of the cluster, but the comparison we show is to the
global values he reported. At higher redshifts, we also compare to
some of the results from \citet{Wylezalek+14}, who measured the \cht\
and \cho\ LF for clusters from the Clusters Around Radio Loud AGN
program \citep[CARLA,][]{Wylezalek+13} in several redshift bins in the
range $1.3 < z < 3.1$. We show two of these in Figures \ref{Fig: mstar
v z} and \ref{Fig: alpha v z}, again where the conversion from \cho\
or \cht\ has a minimal $k$-correction.

\subsubsection{Evolution of the LF}
\label{Sssec: LF evolution}
Figure \ref{Fig: mstar v z} shows the clusters in this work fall into
a larger pattern of passive evolution in $M_H^*$ going out to 
$z\sim2$. Similarly, Figure \ref{Fig: alpha v z} shows very little change
in the faint-end slope over cosmic time when looking only at $\alpha$
over a range of studies. This suggests that the evolution in the
parameters shown in Figure \ref{Fig: covariance_z_comp} is primarily
driven by passive evolution.

\begin{figure}[bthp]
\center{\includegraphics[width=8.5cm]{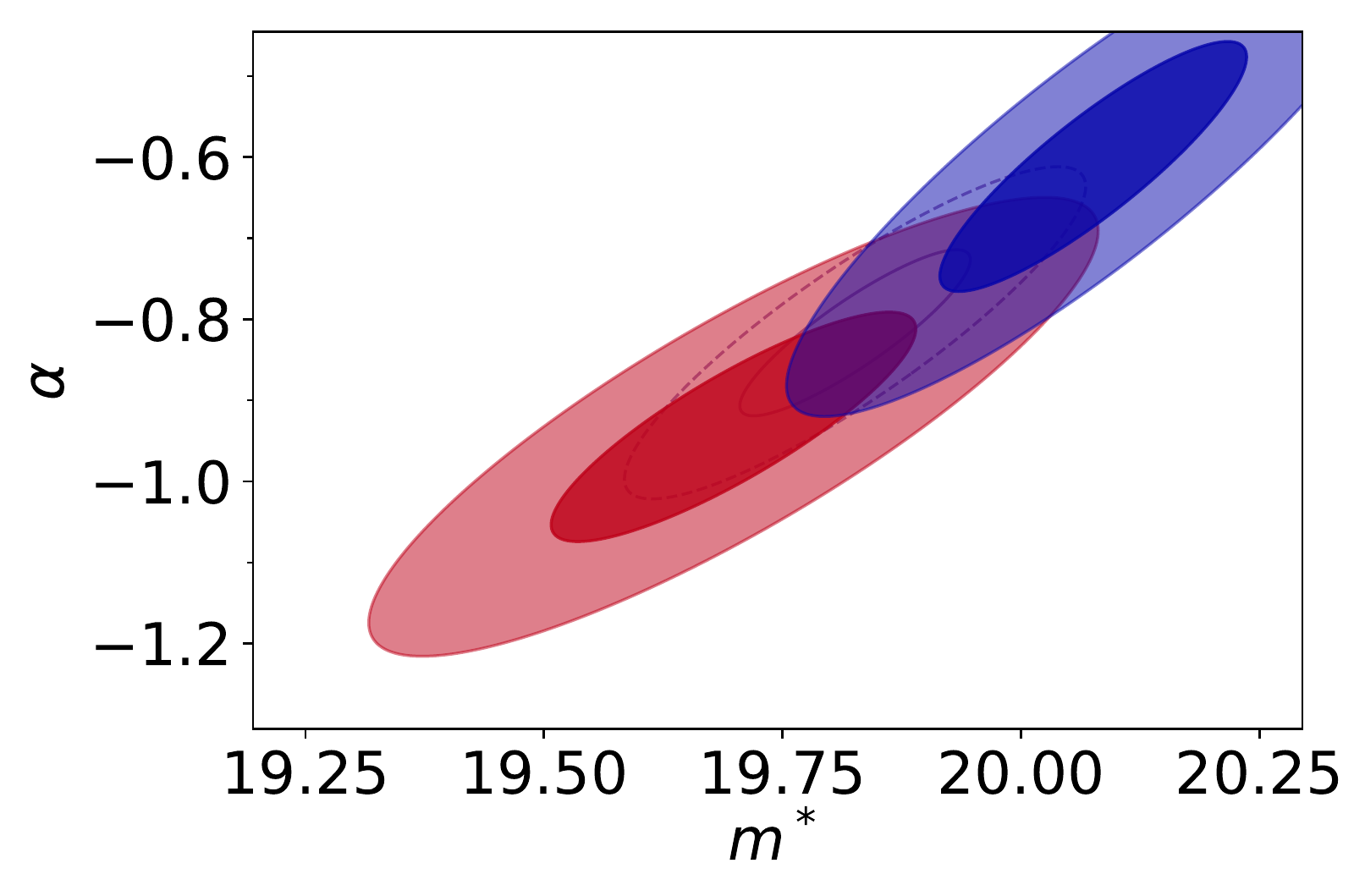}}
\caption{Covariance ellipses showing the $1\sigma$ (dark) and
$2\sigma$ (light) errors and covariances in $\mstar$ and $\alpha$ for
the high-redshift (red) and low-redshift (blue) sub-samples, {\it
  assuming passive evolution}. For comparison, the
whole-sample covariances ellipses from Figure \ref{Fig: covariance_all}
are outlined underneath.}
\label{Fig: evol_corr_ellipses}
\end{figure}

To confirm this, we evolution-correct the galaxies in the high- and
low-redshift sub-samples to $z=1.18$, the mean redshift of the full
sample, assuming passive evolution. We then re-make the LFs and run
the same joint fit as above. The results of this are shown in Figure
\ref{Fig: evol_corr_ellipses}. With passive-evolution `baked-in' to
the fit, the LF parameters are now consistent within two sigma,
supporting the interpretation that these clusters are evolving
passively, consistent with other studies at this redshift.

\section{Conclusions}
\label{Sec: conclusion}
We have presented stellar mass fractions and \cho\ LFs for a sample of \Ncl\
infrared-selected clusters from the \madcows\ catalog. We used
optical and deep mid-IR follow-up data to fit SEDs to objects along
the lines-of-sight to the clusters. This allowed us to more precisely
identify cluster members, measure more thorough stellar masses for the
clusters and measure the faint-end slope of the LF.

The stellar mass fractions we report for these clusters are in good
agreement with previous works, and are consistent with the \citet{GZZ13}
trend line with respect to total mass. For the individual clusters
previously studied in \citet{Decker+19}, the new values of \fstar\
reported here are consistent with---but mostly higher than---the
previous values, with much of the difference being attributable to the
deeper data set we use here.

The composite \cho\ LF we fit for all \Ncl\ clusters has a best-fit
characteristic magnitude and faint-end slope of $\mstar=\mstarall$ and
$\alpha=\alphaall$, respectively. Both are consistent with other works
that have attempted to measure the rest-frame NIR LF for all cluster
members at these redshift ranges. When we split our sample into a
high-redshift bin at $\bar{z}=1.29$ and a low-redshift bin at
$\bar{z}=1.06$ we find that there is a significant evolution in the
best-fit Schechter function parameters, consistent with passive
evolution. This significance is only seen in the covariance ellipse
for $\mstar$ and $\alpha$ jointly. This highlights the need to study
$\mstar$ and $\alpha$ jointly. Comparing to works at other redshifts,
our results are consistent with passive evolution since
$z\sim2$.

In future, follow-up data on more \madcows\ clusters will allow us to
better identify trends with redshift and other cluster parameters. In
addition, deeper infrared data---such as will be attainable from the
next generation of IR space telescopes---will allow us to more
definitively answer questions about the evolution of the faint galaxy
population in clusters at $z > 1$.

\acknowledgments 
The work of T.C., P.E., and D.S. was carried out at the Jet Propulsion
Laboratory, California Institute of Technology, under a contract with
NASA. F.R. acknowledges financial supports provided by NASA through
SAO Award Number SV2- 82023 issued by the Chandra X-Ray Observatory
Center, which is operated by the Smithsonian Astrophysical Observatory
for and on behalf of NASA under contract NAS8-03060.

This work is based in part on observations made with the {\it Spitzer
Space Telescope}, which is operated by the Jet Propulsion Laboratory,
California Institute of Technology under a contract with NASA.
Support for this work was provided by NASA through an award issued by
JPL/Caltech.

This publication makes use of data products from the {\it
Wide-field Infrared Survey Explorer}, which is a joint project of the
University of California, Los Angeles, and the Jet Propulsion
Laboratory/California Institute of Technology, funded by the National
Aeronautics and Space Administration.

This work was based in part on observations obtained at the Gemini
Observatory, which is operated by the Association of Universities for
Research in Astronomy, Inc., under a cooperative agreement with the
NSF on behalf of the Gemini partnership: the National Science
Foundation (United States), the National Research Council (Canada),
CONICYT (Chile), Ministerio de Ciencia, Tecnolog\'{i}a e
Innovaci\'{o}n Productiva (Argentina), and Minist\'{e}rio da
Ci\^{e}ncia, Tecnologia e Inova\c{c}\~{a}o (Brazil).

Funding for this program is provided by NASA through the NASA
Astrophysical Data Analysis Program, award 80NSSC19K0582. Support for
this work was provided by the National Aeronautics and Space
Administration through Chandra Award Number GO7-18123A issued by the
Chandra X-ray Center, which is operated by the Smithsonian
Astrophysical Observatory for and on behalf of the National
Aeronautics Space Administration under contract NAS8-03060. Based on
observations with the NASA/ESA Hubble Space Telescope obtained at the
Space Telescope Science Institute, which is operated by the
Association of Universities for Research in Astronomy, Incorporated,
under NASA contract NAS5-26555. Support for program number
HST-GO-14456 was provided through a grant from the STScI under NASA
contract NAS5-26555.

%\newpage

%\bibliographystyle{aasjournal}
\bibliographystyle{yahapj}
\bibliography{allrefs}

\end{document}